\documentclass[preprint2]{aastex}

\usepackage{graphicx}
\DeclareGraphicsExtensions{.ps}  
\shorttitle{\indent \def Oscillations above light bridges} \shortauthors{Zhang et al.}

\begin{document}

\title{Surge-like oscillations above sunspot light bridges driven by magnetoacoustic shocks}

\author{Jingwen Zhang\altaffilmark{1}, Hui Tian\altaffilmark{1}, Jiansen He\altaffilmark{1}, Linghua Wang\altaffilmark{1}}
\altaffiltext{1}{School of Earth and Space Sciences, Peking University, 100871 Beijing, China; huitian@pku.edu.cn}
%\altaffiltext{2}{College of Science, George Mason University, Fairfax, VA 22030, USA}

\begin{abstract}
High-resolution observations of the solar chromosphere and transition region often reveal surge-like oscillatory activities above sunspot light bridges. These oscillations are often interpreted as intermittent plasma jets produced by quasi-periodic magnetic reconnection. We have analyzed the oscillations above a light bridge in a sunspot using data taken by the Interface Region Imaging Spectrograph (IRIS). The chromospheric 2796\AA{}~images show surge-like activities above the entire light bridge at any time, forming an oscillating wall. Within the wall we often see that the Mg~{\sc{ii}}~k 2796.35\AA{}~line core first experiences a large blueshift, and then gradually decreases to zero shift before increasing to a red shift of comparable magnitude. Such a behavior suggests that the oscillations are highly nonlinear and likely related to shocks. In the 1400\AA{}~passband which samples emission mainly from the Si~{\sc{iv}}~ion, the most prominent feature is a bright oscillatory front ahead of the surges. We find a positive correlation between the acceleration and maximum velocity of the moving front, which is consistent with numerical simulations of upward propagating slow-mode shock waves. The Si~{\sc{iv}} 1402.77\AA{}~line profile is generally enhanced and broadened in the bright front, which might be caused by turbulence generated through compression or by the shocks. These results, together with the fact that the oscillation period stays almost unchanged over a long duration, lead us to propose that the surge-like oscillations above light bridges are caused by shocked p-mode waves leaked from the underlying photosphere. 
\end{abstract}


\section{Introduction}
Light bridges (LBs) are one of the most prominent features in some sunspots. They are structures extending into dark umbrae and sometimes even disintegrating sunspots completely. Previous studies show that the magnetic field of LBs has lower field strength and more horizontal inclination than that of neighboring umbral regions \citep[e.g.,][]{Leka1997,Shimizu2009,Toriumi2015a,Yuan2016}. \cite{Jurcak2006} came up with a canopy structure of LBs magnetic field, that the magnetic field lines on both sides tilt inward and meet each other in the middle. The real magnetic field structure of LBs must be much more complicated\citep{Toriumi2015a}. Hence the discrepancy between the magnetic field of LBs and surrounding umbral areas can result in many interesting phenomena above LBs. 

One of these interesting phenomena is the persistent plasma oscillations in the chromosphere and transition region (TR) above light bridges. The earliest observation of oscillations above LBs was reported as recurrent surges in the H$\alpha$ passband originating from Ellermann Bombs (EBs) \citep{Roy1973}. \cite{Asai2001} measured the mean velocity, maximum length and mean life time of the H$\alpha$ surges above LBs to be 50~km~s$^{-1}$, 20~Mm and 10 minutes, respectively. They also detected a U-shaped counterpart in the 171\AA{} passband of the Transition Region and Coronal Explorer \citep[TRACE,][]{Handy1999} at the same time and location. This structure shows no emission in soft X-ray passband. With unprecedented high resolution of modern space and ground-based telescopes, more details of these surges have been revealed recently. For instance, \cite{Shimizu2009} observed similar chromospheric ejections with Ca~{\sc{ii}} H imaging by Hinode \citep{Kosugi2007}. A chain of ejections seemed to be launched from along the edge of the LB intermittently and recurrently, possibly tracing the fan-shaped magnetic field lines. The detection of strong current density in the middle of the LB led them to conjecture the following scenario: the current-carrying magnetic flux ropes trapped below a cusp-shaped magnetic structure along the LB reconnected with pre-existing upright umbral magnetic fields. This observational result seems to be different from that of \cite{Toriumi2015a}, who detected strong current density at the edge rather than in the middle of the LB. Based on both observations and numerical simulations, \cite{Toriumi2015a,Toriumi2015b} suggested that chromospheric brightenings and dark surge ejections in a LB are driven by magnetoconvective evolutions that have convective upflows in the center and downflows at the edge. Through the convection, horizontal magnetic fields are carried from the solar interior to the surface and reconnect with surrounding umbral magnetic fields continuously. \cite{Robustini2016} also reported fan-shaped jet-like activities above LBs and considered magnetic reconnection as the driver. Similar jet-like activities have also been identified from a LB by \cite{Louis2014}, and again reconnection was proposed as the driver of these activities. More recently, through observations of the Interface Region Imaging Spectrograph \citep[IRIS,][]{DePontieu2014}, \cite{Bharti2015} and \cite{Yang2015} independently identified a bright oscillatory front ahead of the H$\alpha$ surges above the LB within a sunspot. \cite{Bharti2015} interpreted the bright front as heating to TR temperatures by either shock front or compression. \cite{Yang2015} named such oscillations light wall oscillations and conjectured that they are driven by leaked p-mode waves from below the photosphere, based on a joint observation between IRIS and the Chinese 1-m New Vacuum Solar Telescope \citep[NVST,][] {Liu2014}. In addition, \cite{Hou2016a} and \cite{Yang2016} reported that external disturbances such as flares and falling materials could enhance the amplitude of the light wall oscillations.

It appears that most previous studies attributed these surge-like oscillations to magnetic reconnection. Some of these surges are preceded by obvious brightennings at the LBs, extend up to 40 Mm, and propagate with a speed up to 100~km~s$^{-1}$. These observational facts do support the reconnection interpretation. However, in many observations these relatively long surges seem to coexist with shorter and constantly occurring surges. The latter appears to reveal coordinated behaviors between neighboring surges repeatedly launched from along LBs. As already noticed by \cite{Bharti2015}, magnetic reconnection may not easily explain the coordinated behaviors of these surge-like oscillations. Through analysis of a dataset taken by IRIS, we find that these oscillations may be better explained as a consequence of shock waves generated from the upward propagating p-mode. 

\section{Observations}

We analyze the 8-step-raster observations made from 22:35 UT on 2015 September 29 to 03:32 UT on the next day. The pointing coordinate was (666$^{\prime\prime}$, --375$^{\prime\prime}$), targeting NOAA active region (AR) 12422. The spatial pixel size was 0$^{\prime\prime}$.167. The field of view (FOV) of the spectral observation was 8$^{\prime\prime}$$\times$120$^{\prime\prime}$, as one raster scan consisted of eight 1$^{\prime\prime}$ steps. The step cadence was 5.3 s and the raster cadence of spectral observation in both the near-ultraviolet (NUV, 2783-2834\AA{}) and far-ultraviolet (FUV, 1332-1358\AA{} and 1390-1406\AA{}) wavelength bands was 42 s. Slit-jaw images (SJIs) in the filters of 1400\AA{}, 1330\AA{} and 2796\AA{} were taken alternatively and the cadence for each filter was 21 s. The FOV of SJIs was 120$^{\prime\prime}$$\times$120$^{\prime\prime}$. We used the level 2 data, where dark current subtraction, flat field, geometrical and orbital variation corrections have been applied \citep{DePontieu2014}. Fiducial marks on the IRIS slit were used to coalign the SJIs with the spectral data. 

We use mainly the Si~{\sc{iv}} 1402.77\AA{} and Mg~{\sc{ii}} k 2796.35\AA{} lines formed in the middle TR (formation temperature $\sim$0.1 MK) and chromosphere (formation temperature $\sim$0.01 MK) respectively in this study. For absolute wavelength calibration in the FUV wavelength band, we assumed that the optically thin Fe~{\sc{ii}} 1392.817\AA{} line has no net Doppler shift on average. This assumption can be justified since the cold Fe~{\sc{ii}} line has a very small intrinsic velocity \citep[e.g.,][]{Tian2016}. SJI images of the 1400\AA{} and 2796\AA{} passbands are also analyzed. The former contains mainly two Si~{\sc{iv}} lines and the UV continuum emission. The latter samples mainly the Mg~{\sc{ii}} k line emission. Figure~\ref{fig.1} shows a snapshot of the IRIS observation. 

\section{Data analysis and results}

\begin{figure*}
\centering {\includegraphics[width=\textwidth]{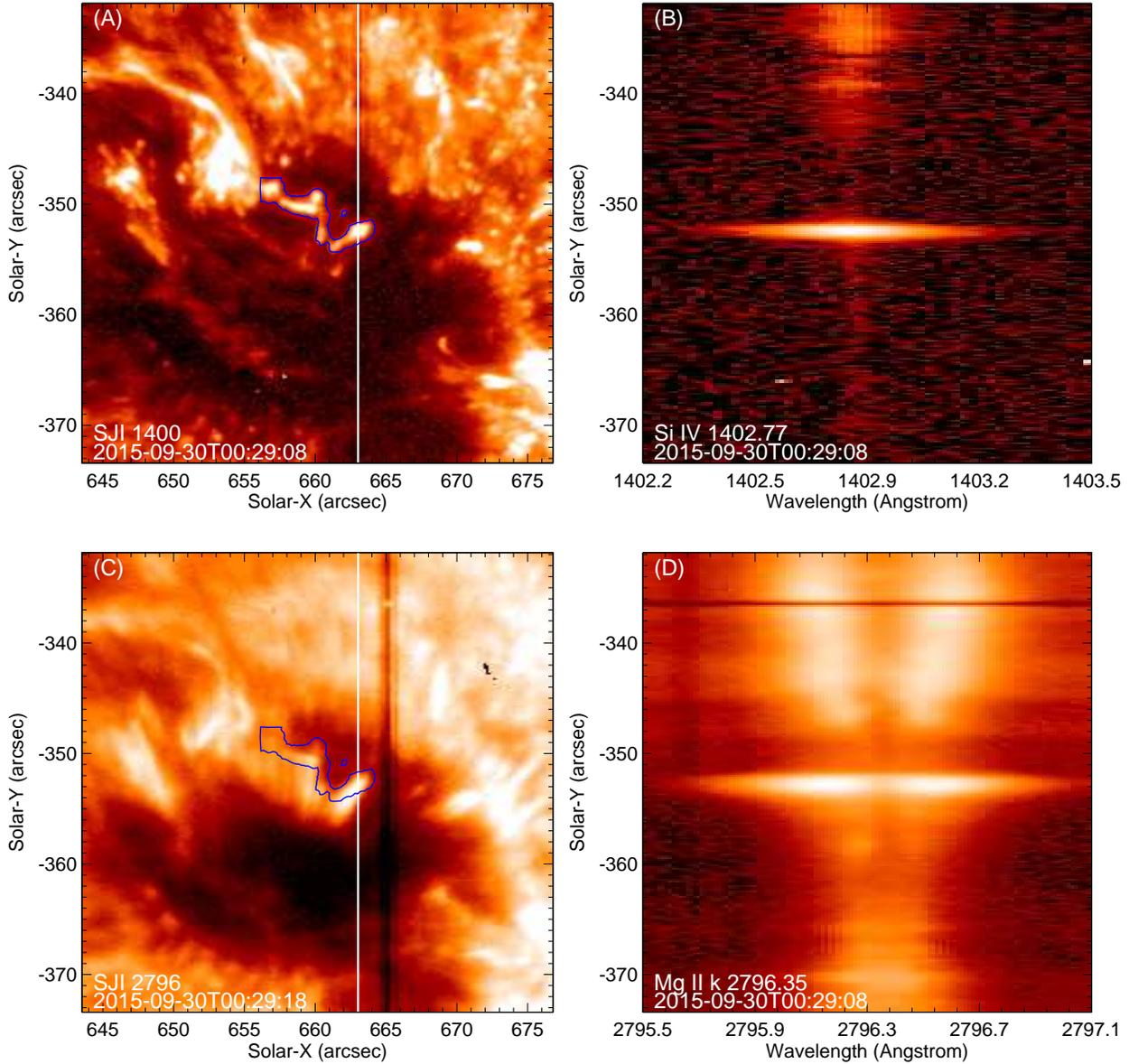}} \caption{ (A) \& (C) IRIS SJI images in the 1400\AA{} and 2796\AA{} passbands taken respectively at 00:29:08 UT and 00:29:18 UT. (B) \& (D) Si~{\sc{iv}} 1402.77\AA{} and Mg~{\sc{ii}} k 2796.35\AA{} line spectra along the white lines in (A) \& (C) at 00:29:08 UT.  The bright front in the 1400\AA{} image is marked as the blue contour in (A) \& (C). The dark horizontal lines in the spectral images are caused by the fiducial marks on the slit. An animation showing the IRIS observation is available online.} \label{fig.1}
\end{figure*}

From Figure~\ref{fig.1} and the online animation, we can see surge-like oscillations during the entire observation period in the 2796\AA{} passband. They should be the same phenomenon as the recurrent surges or light wall previously detected in the H$\alpha$ and Ca~{\sc{ii}} images \citep[e.g.,][]{Asai2001,Shimizu2009,Yang2015}. In the 1400\AA{} images, we generally see an oscillating bright front and the nearly stationary light bridge which seem to encompass the wall. The bright front keeps going up and down, possibly accompanied by intensity enhancement along the light bridge. To examine the relationship between oscillations in the two passbands, we outline the bright front in the 1400\AA{} image as the contour in the 2796\AA{} image. The bright front in the 1400\AA{} image is located exactly on top of the bright hazy region (the surges or light wall) in the 2796\AA{} image, suggesting that the oscillations in the two passbands are manifestations of the same oscillations. The different appearance and locations might be caused by the different responses to temperature in the two passbands. It is likely that the surges are mainly chromospheric plasma, which explains why they are usually observed in chromospheric passbands such as H$\alpha$, Ca~{\sc{ii}} H and Mg~{\sc{ii}} k 2796\AA{}. As these cool surges hit the upper layers of the atmosphere, compression effect takes effect and the plasma ahead of the surges are heated to TR temperatures, revealing as a bright dynamic front in the 1400\AA{} images. 

\begin{figure*}
\centering {\includegraphics[width=\textwidth]{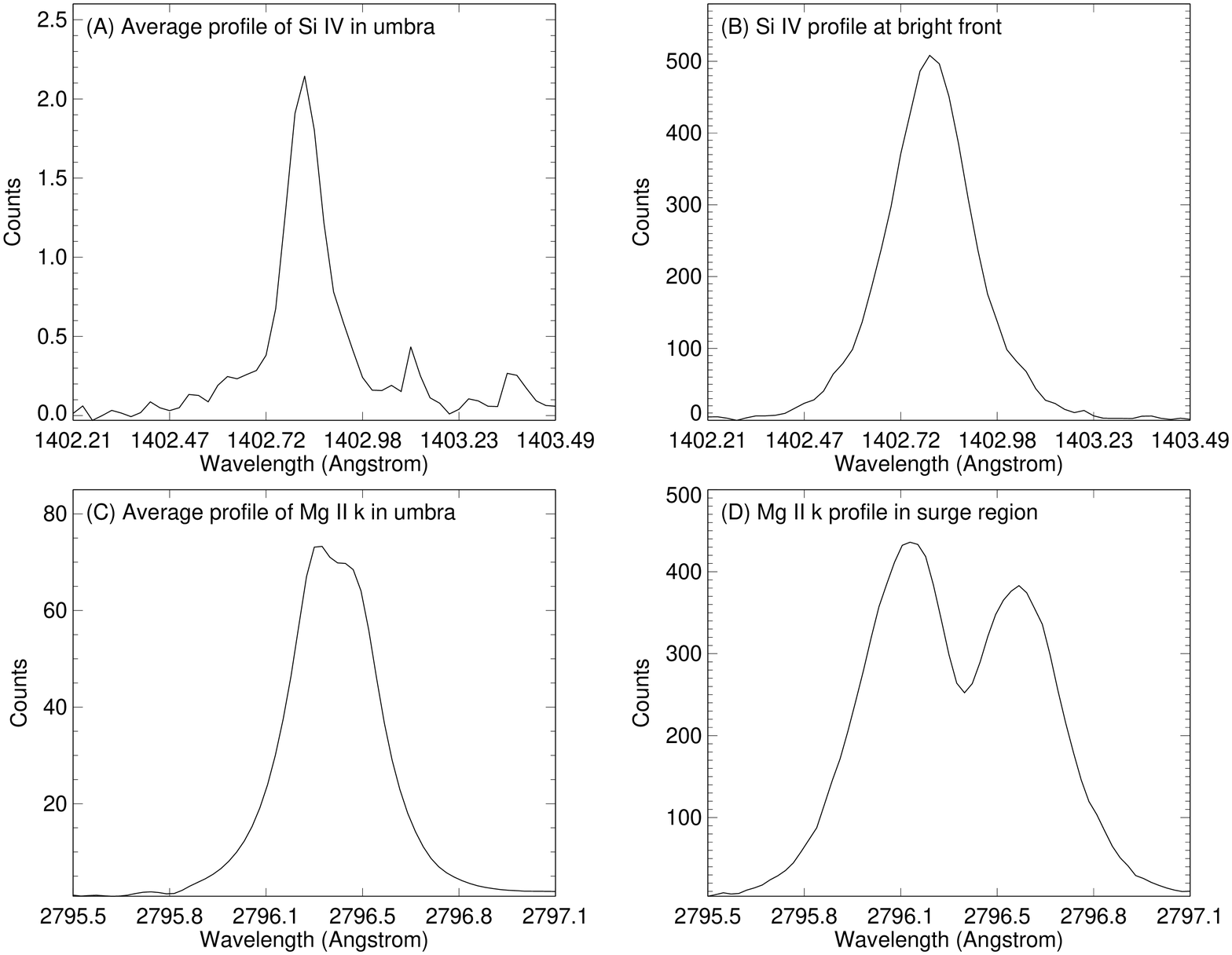}} \caption{(A) \& (C) Average profiles of the Si~{\sc{iv}} 1402.77\AA{} and Mg~{\sc{ii}} k 2796.35\AA{} lines in the umbral region (Solar-Y=-373$^{\prime\prime}$ $\sim$ -357$^{\prime\prime}$). (B) Profile of the Si~{\sc{iv}} 1402.77\AA{} line at the bright front (Solar-Y=-353$^{\prime\prime}$ observed at 00:29:08 UT. (D) Profile of the Mg~{\sc{ii}} k 2796.35\AA{} line within the light wall (Solar-Y=-353$^{\prime\prime}$ observed at 00:29:08 UT. }
\label{fig.2}
\end{figure*}

Figure~\ref{fig.1}(B) \& (D) show the spectral images of the Si~{\sc{iv}} 1402.77\AA{} and Mg~{\sc{ii}} k 2796.35\AA{} lines. At 00:29:08 UT, the slit happens to catch the bright front (Solar-Y is around -353$^{\prime\prime}$). Figure~\ref{fig.2}(B) shows a typical profile of the Si~{\sc{iv}} 1402.77\AA{} line in the bright front and Figure~\ref{fig.2}(D) shows a typical profile of the Mg~{\sc{ii}} k 2796.35\AA{} line within the light wall at this time. For the purpose of comparison, Figures~\ref{fig.2}(A) \& (C) show respectively the average profiles of the Si~{\sc{iv}} 1402.77\AA{} and Mg~{\sc{ii}} k 2796.35\AA{} lines in the umbral region. The Si~{\sc{iv}} 1402.77\AA{} line profile reveals an obvious intensity enhancement in the bright front, implying that the Si~{\sc{iv}} 1402.77\AA{} line emission contributes a lot to the bright front observed in the 1400\AA{} passband. The slit also overlaps with part of the oscillating surges in the SJI 2796\AA{} passband. Although Mg~{\sc{ii}} k 2796.35\AA{} line profile usually has a reversal at the line core, its opacity reduces in sunspots and the profile is close to Gaussian distribution because of the tenuous atmosphere above sunspots (Figure~\ref{fig.2}(C)) \citep{Tian2014a}. However, from Figure~\ref{fig.1}(D) and Figure~\ref{fig.2}(D), we can see that the Mg~{\sc{ii}} k 2796.35\AA{} line profile has a darker core compared to the near wings at locations of the surges. Obviously, the Mg~{\sc{ii}} k 2796.35\AA{} line is optically thicker in the surges above the light bridge than in the surrounding umbral region. 

\begin{figure*}
\centering {\includegraphics[width=\textwidth]{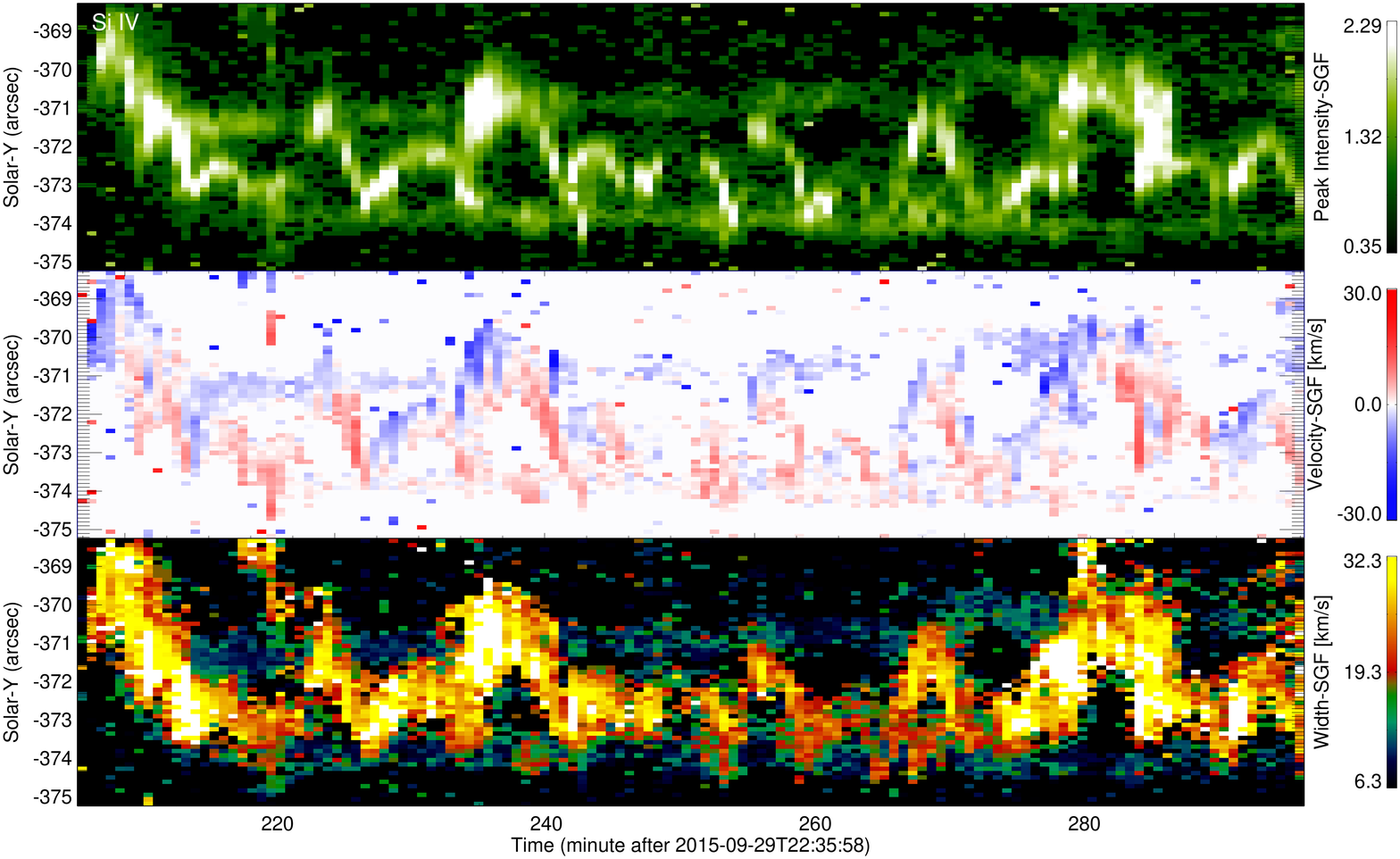}} \caption{ Temporal evolution of the Si~{\sc{iv}} 1402.77\AA{} SGF parameters within a segment of the slit. From top to bottom:  Peak Intensity, Doppler Shift and Doppler Width. Only 96-minute data are shown here. }
\label{fig.3}
\end{figure*}

The Si~{\sc{iv}} 1402.77\AA{} line is generally optically thin, we thus apply a single Gaussian fit (SGF) to all Si~{\sc{iv}} 1402.77\AA{} line profiles along the slit. The temporal evolution of the SGF parameters are shown in Figure~\ref{fig.3}. In the images of peak intensity, Doppler velocity and Doppler width, similar oscillation patterns appear after wiping out noises. Ascending and descending motions of the surge-related oscillating front are clearly present in the images. We can see that the ascending and descending phases are dominated by  blue and red shifts with a magnitude of a few~km~s$^{-1}$, respectively. One should be aware that these velocities are just the line-of-sight (LOS) projection of the real velocities. The bottom panel clearly shows that the Si~{\sc{iv}} line profiles are often greatly broadened in the bright front. The typical value of the Dopper width is about 30~km~s$^{-1}$. The nonthermal width is of the same order, if we assume a thermal broadening of 6.8~km~s$^{-1}$ and an instrumental broadening of 4.1~km~s$^{-1}$ \citep[see the online supplementary materials of][]{Tian2014b}.

\begin{figure*}
\centering {\includegraphics[width=0.9\textwidth]{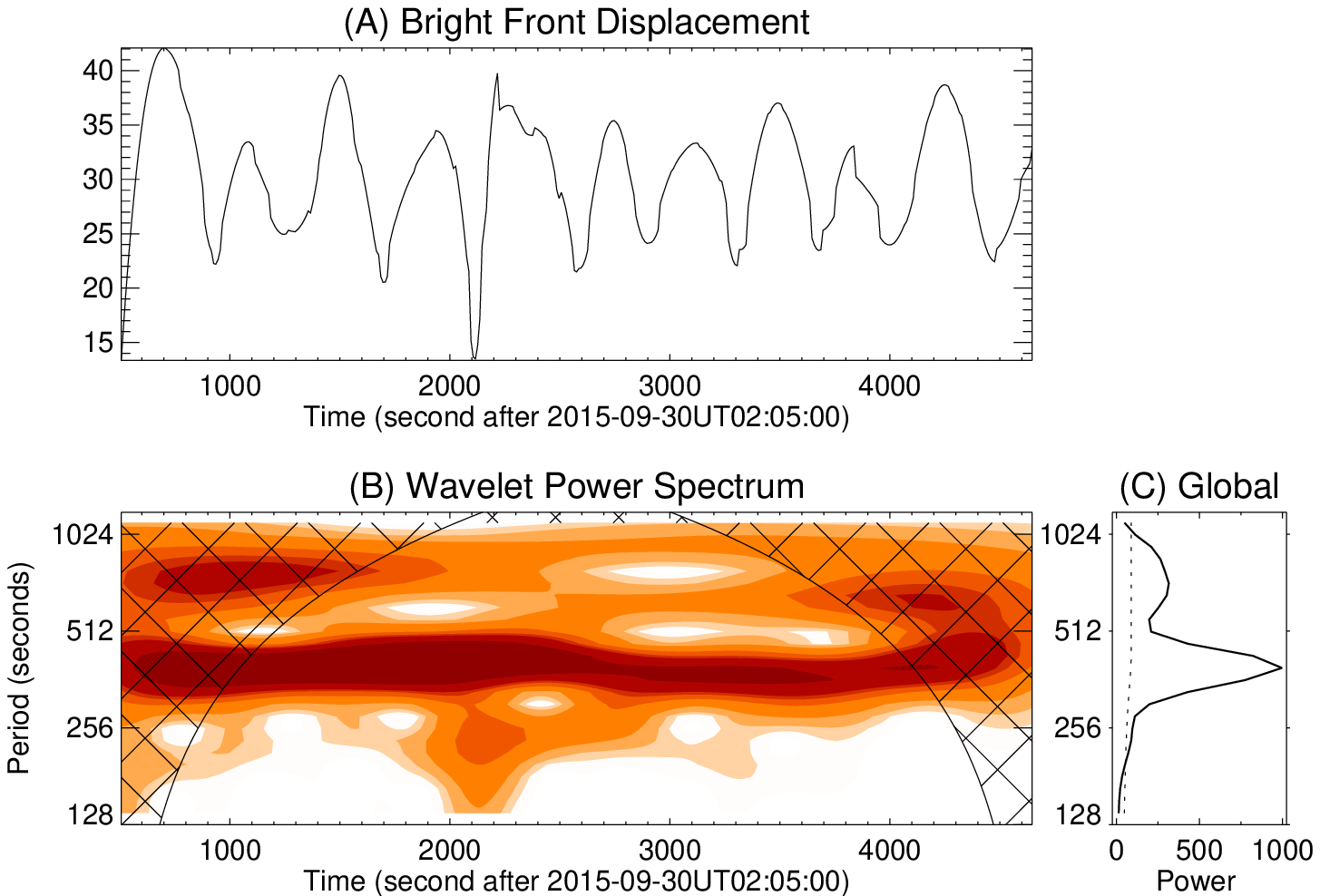}} \caption{ (A) Time series of the displacement of the bright front derived from the top panel of Figure~\ref{fig.3}. The displacement is expressed in the unit of spatial pixel (0$^{\prime\prime}$.167). The wavelet spectrum and global wavelet for the time series are shown in (B) and (C), respectively. Dark color represents strong power.} \label{fig.4}
\end{figure*}

To determine the period of the oscillation, we first obtain a time series of the displacement of the bright front by manually clicking on the top panel of Figure~\ref{fig.3}, then remove a background trend by subtracting a 400-second boxcar smoothing of the time series from the original one. The wavelet analysis result shown in Figure~\ref{fig.4} reveals a power peaked sharply around 6.5 minutes. The period appears constant over almost an hour. 

\begin{figure*}
\centering {\includegraphics[width=0.9\textwidth]{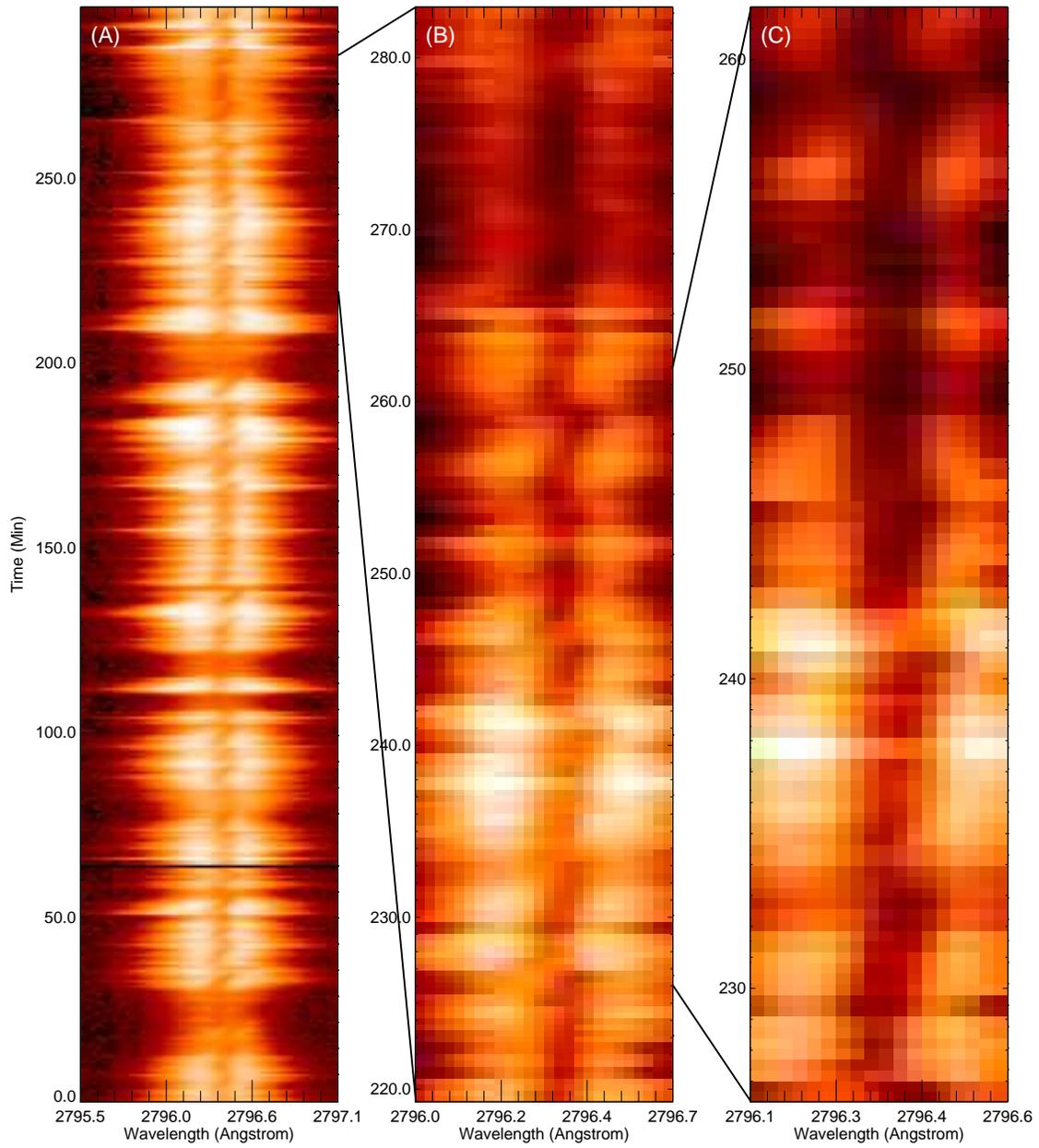}} \caption{ (A) Wavelength-time diagrams for the Mg~{\sc{ii}} k 2796.35\AA{} line at the slit position of Solar-Y = -353$^{\prime\prime}$. (B) \& (C) Same as (A) but for shorter time ranges.} \label{fig.5}
\end{figure*}

As we mentioned above, the line core of Mg~{\sc{ii}} k 2796.35\AA{} appears darker than the near wings in the surges. Temporal evolution of the Mg~{\sc{ii}} k line profiles at Solar-Y = -353$^{\prime\prime}$ are shown in Figure~\ref{fig.5}. According to the Doppler effect, the variation of wavelength is equivalent to the change of velocity in the line of sight. The dark line core clearly repeats the following behavior: a fast impulsive blueward excursion followed by a gradual redward excursion. That is to say, the plasma that radiate Mg~{\sc{ii}} k emission first experiences a sudden upward acceleration at low height, then moves upward with decreasing velocity, and finally drops down with increasing speed. Such behavior is repeated during the entire observation period.

\begin{figure*}
\centering {\includegraphics[width=0.9\textwidth]{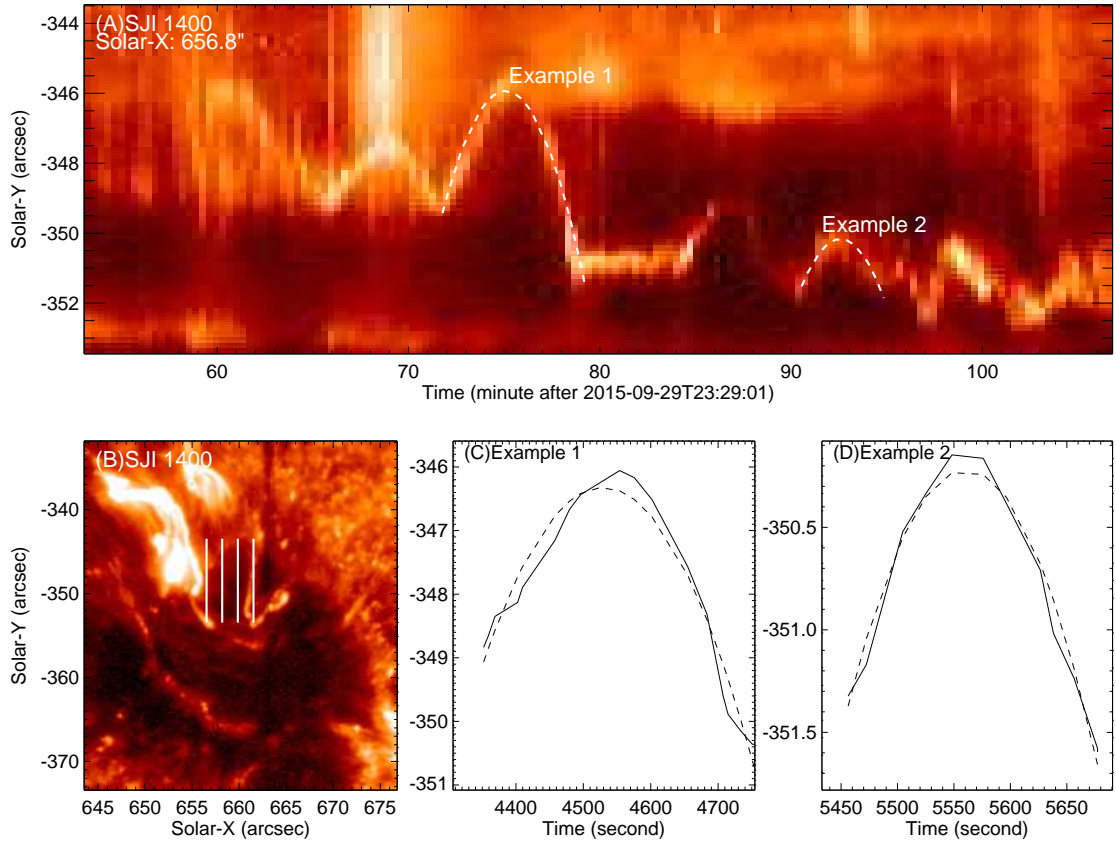}} \caption{ Parabolic fitting of the path of the oscillating front. (B) An SJI image of the 1400\AA{} passband. The four white lines mark the cuts for which we produce space-time diagrams. (A) Part of the space-time diagrams. Two examples of parabolic trajectories of the bright front are marked using the dashed lines. (C) \& (D) Parabolic fitting of the two trajectories marked in (A).} \label{fig.6}
\end{figure*}

The up-and-down motion of the bright front in the 1400\AA{} images appear to follow a parabolic path. In Figure~\ref{fig.6}(A) we show part of the space-time diagram for one of the vertical cuts marked in Figure~\ref{fig.6}(B). Parabolic trajectories are clearly seen in the space-time diagram. We have produced the space-time diagrams for all the four vertical cuts marked in Figure~\ref{fig.6}(B). From the four diagrams we have identified 98 well-defined and well-isolated trajectories and performed a parabolic fitting of them. Figure~\ref{fig.6}(C) \& (D) show  two examples of the fitting. 

\begin{figure*}
\centering {\includegraphics[width=0.7\textwidth]{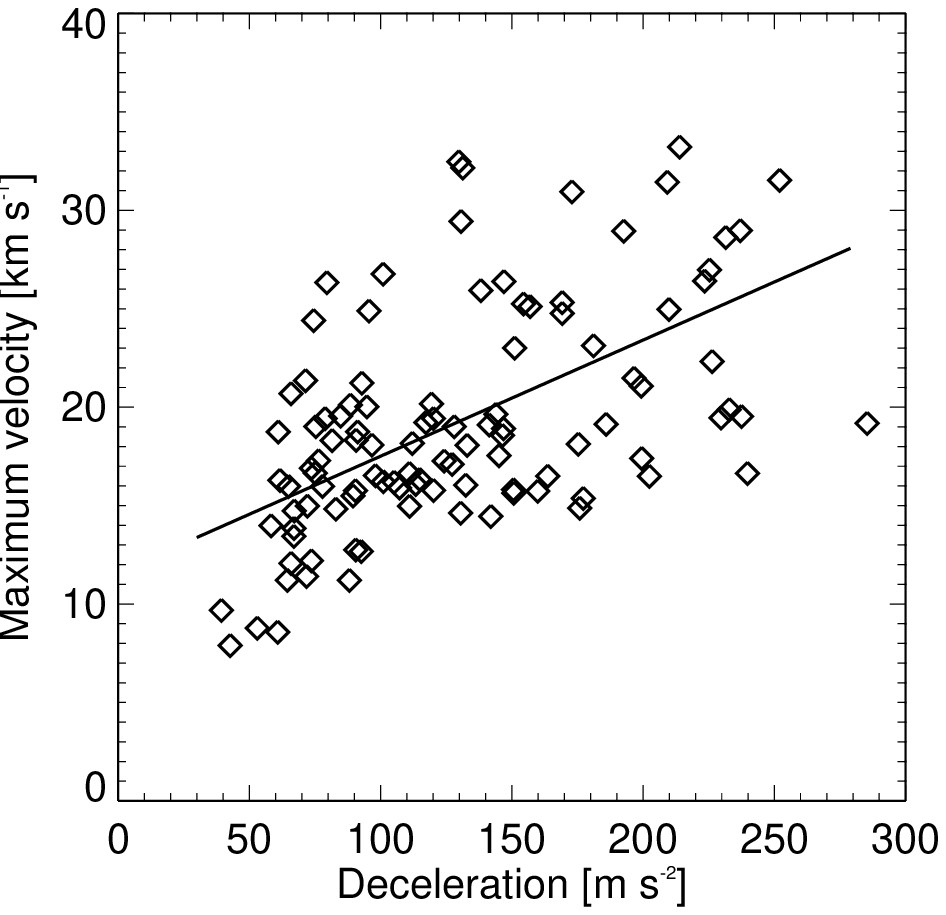}} \caption{ Scatter plots of the relationship between maximum velocity and deceleration obtained from a parabolic fitting of the bright front trajectories. The solid line represent the linear fitting.}  \label{fig.7}
\end{figure*}

From the fitting we have derived the maximum velocity and deceleration of the motion for each parabola. Figure~\ref{fig.7} shows the scatter plots of the relationship between maximum velocity and deceleration. We find that the maximum velocity is generally between 10  and 30~km~s$^{-1}$. The deceleration is mostly in the range of 50 and 250~m~s$^{-2}$, which is smaller than the gravitational deceleration on the Sun. Figure~\ref{fig.7} reveals a positive correlation between the maximum velocity and deceleration.

\begin{figure*}
\centering {\includegraphics[width=0.9\textwidth]{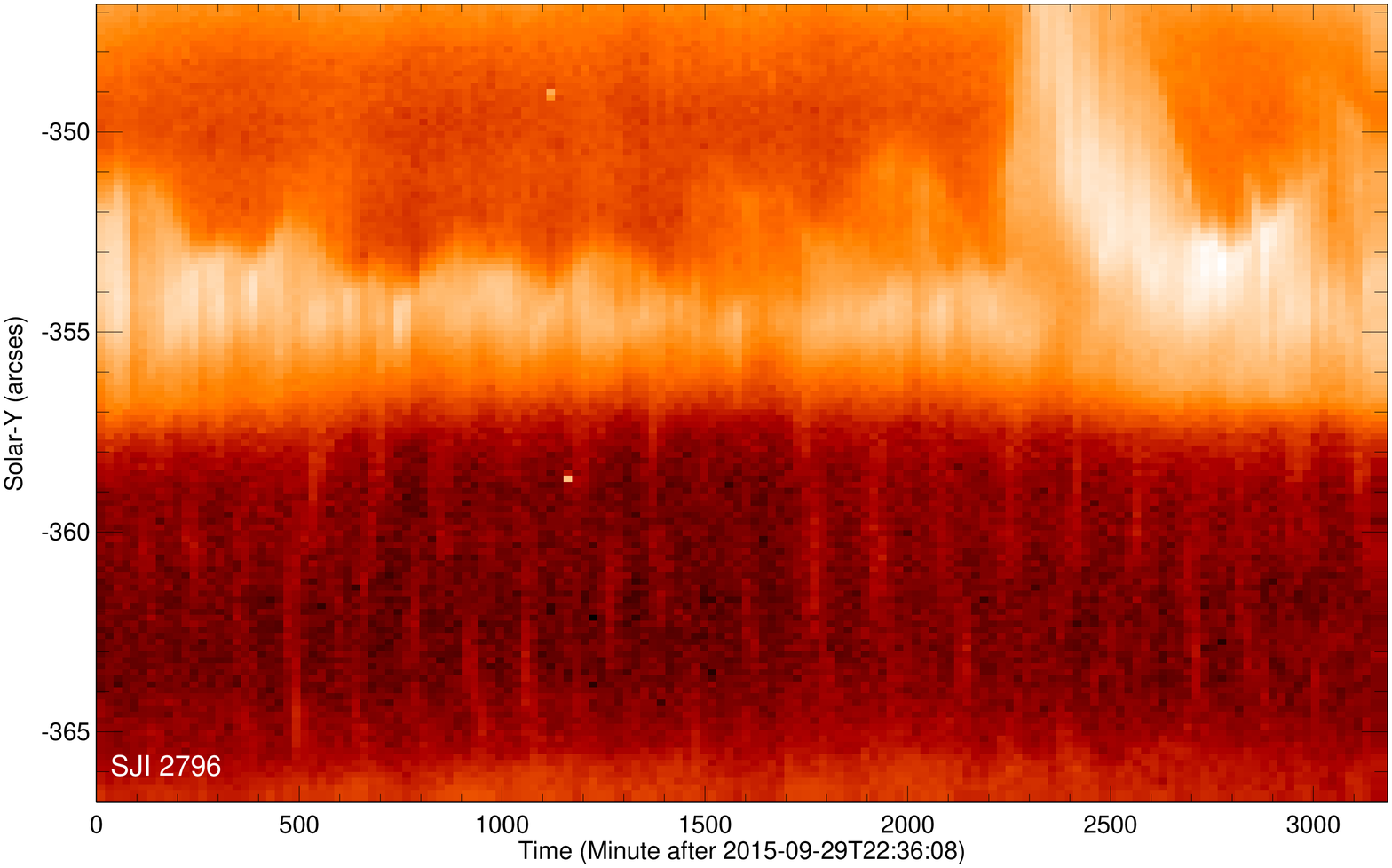}} \caption{ Space time diagram of the 2796\AA{} SJI images at Solar-X= 660.16$^{\prime\prime}$. The upper part shows the surge-like activities/light wall oscillation. The lower part shows umbral oscillations. } \label{fig.8}
\end{figure*}
 
Figure~\ref{fig.8} presents the space time diagram of the 2796\AA{} SJI images at Solar-X = 660.16$^{\prime\prime}$. It reveals not only light bridge oscillations but also the surrounding umbral oscillations. There appears to be no chronological relation between them. The different periods of the two types of oscillations are also evident from this figure: umbral oscillations are dominated by a period around 3 minutes, while the oscillations above light bridges show a longer period. 

\section{Discussion and Conclusion}
Previous observations of the sunspot's chromosphere often reveal surge-like oscillatory activities above light bridges. These oscillations were usually interpreted as reconnection driven jets from the light bridges. More recently, IRIS observations have demonstrated that these surges often are led by a bright front at transition region temperatures. In this paper, we have confirmed that the chromospheric surges and transition region front are manifestations of the same quasi-periodic process. 

However, our observation and analysis results do not support the interpretation of these light bridge oscillations as reconnection driven jets. First, the 2796\AA{}~images show surge-like activities above the entire light bridge at any time. This would require the occurrence of reconnection everywhere on the light bridge at all times if reconnection is the driving mechanism. This scenario appears not likely as reconnection at the light bridge usually occurs between the background fields and small-scale emerging fluxes. And the latter definitely can not appear at all locations on the light bridges and at any time \citep[e.g.,][]{Louis2015}. Second, the bright oscillating front in the 1400\AA{}~images shows well-coordinated behaviors between neighboring pixels, leading to only one bright front. If magnetic reconnection is the driving mechanism of the oscillations, the surges should be triggered randomly in time and in space. In that case the bright front should be a broken line rather than a continuous line. Third, our wavelet analysis has revealed a nearly stationary oscillation period around 6.5 minutes. Reconnection may occur quasi-periodically. But in that case one should expect a large scatter in the observed period. Fourth, the maximum velocity of the bright front is generally in the range of 10$\sim$30~km~s$^{-1}$, which is much lower than local Alfv\'en velocity in the transition region. As we know, reconnection-driven jets should move at a velocity of the order of the local Alfv\'en speed. 

On the other hand, the observational facts mentioned above can be easily explained by slow-mode shock waves generated as the p-mode leaks from the photosphere to the chromosphere. This scenario has been previously proposed to explain the dynamic fibrils in plage regions \citep{DePontieu2004,DePontieu2007,Hansteen2006}  and umbral oscillations \citep[e.g.,][]{Rouppe2003,Centeno2006,Tian2014a}. Indeed, the morphology of the surge-like activities above light bridges is highly similar to that of dynamic fibrils in chromospheric images. P-mode oscillations are ubiquitous in the solar atmosphere and they never stop, which explains the wall morphology above the light bridge at any time. P-mode oscillations are also known to be coherent over a certain spatial range. For instance, in sunspots we often see a clear wave front, which is believed to be related to shock waves propagating along different field lines \citep[e.g.,][]{Rouppe2013,Chae2014,Tian2014a,Madsen2015}. The bright oscillating front in our observation could be related to a similar process. The dominant period of the leaked p-mode depends on the inclination of the magnetic field lines. In the strong-field sunspot, magnetic field line inclination is unexpected to change dramatically over time. Thus, the leaked p-mode should have a more or less constant period. Finally, the maximum velocities of the surges or bright front in our observation are in the range of 10$\sim$30~km~s$^{-1}$, which is also comparable to those of dynamic fibrils \citep{DePontieu2007,Hansteen2006}. 

Spectroscopic observations have also revealed signatures of shock waves. Temporal evolution of the Mg~{\sc{ii}} k 2796.35\AA{} line profiles reveals a repeated pattern of the following behavior: the dark line core first experiences a sudden impulsive blueward excursion and then a gradual shift to the red side. Such behavior can be interpreted as the radiative plasma being accelerated abruptly, moving upward with decreasing velocity and then dropping down. In principle the initial velocity pulse can be produced by shocks or magnetic reconnection. And shock waves may also develop from slow-mode waves generated by reconnection \citep{Song2017}. However, as mentioned above, magnetic reconnection appears to have difficulty in explaining the nearly constant period of the oscillations and the coordinated behavior of a chain of surges. Similar spectral behavior has been frequently observed in sunspot oscillations and cited as evidence of shock waves \citep[e.g.,][]{Thomas1984,Rouppe2003,Centeno2006,Felipe2010,Tian2014a}.

\cite{Robustini2016} has noticed that the paths of some H$\alpha$ surges above light bridges can be approximated by parabolas. Here we further demonstrate that the motion of the bright moving front in the 1400\AA{} images also follow a parabolic path. This behavior is consistent with the spectral shift of the Mg~{\sc{ii}} k 2796.35\AA{} lines mentioned above, as a parabolic motion implies a constant acceleration and linearly varied velocities. In addition, we find a positive correlation between the maximum velocity and acceleration/deceleration. A similar correlation has been previously found for dynamic fibrils and umbral oscillations as well \citep{DePontieu2007,Hansteen2006,Tian2014a}. \cite{Hansteen2006} and \cite{DePontieu2007} simulated upward propagating magnetoacoustic shock waves and reproduced the observational behavior of dynamic fibrils. They found that the highly dynamic chromospheric shock waves cause significant up- and downward excursions of the upper chromosphere. These simulations predict a positive correlation between the deceleration and the maximum velocity. Given the similar morphology and many similar properties between the dynamic fibrils and the light bridge oscillations studied in this paper, we believe that the same process might also be the cause of the surge-like oscillations above light bridges. The finding of similar light wall oscillations outside sunspots \citep{Hou2016b} also strengthens this conclusion. 

Our observation clearly shows that the leading edge of the 2796\AA{}~surges is hotter than the body of the surges. The leading edge exhibits as a bright oscillating front in the 1400\AA{} images. The Si~{\sc{iv}} 1402.77\AA{}~line profile is clearly enhanced in the bright front, which might be caused by compression of the upward moving surges with the overlying plasma. The greatly enhanced nonthermal line width of Si~{\sc{iv}} 1402.77\AA{}~might be caused by micro-turbulence generated through compression or by the shocks. 

Our analysis also reveals different periods of the umbral oscillations and the light bridge oscillations. \cite{Yuan2014} found the same result and they attributed it to the departure of the magnetic field lines from the vertical over the bridge. The 6.5-minute period in our observation could be caused by the more inclined magnetic field lines around the light bridge as well. It is known that the inclination of the magnetic field lines lowers the acoustic cut-off frequency, thus allowing longer-period p-mode waves to leak upward \citep[e.g.,][]{Sych2014,Jess2016,Sobotka2013,Madsen2015}. We notice that \cite{Yang2015} reported a 3.9-minute period of the light bridge oscillations in their dataset. The different periods observed in different light bridges are likely related to different inclination angles of the magnetic fields lines. In principle it is possible to estimate the inclination angle of a field-aligned surge as we can measure the LOS component and plane of sky (POS) component of the surge velocity respectively from our spectroscopic and imaging observations. However, in our case the LOS velocity component is generally a few km~s$^{-1}$ and its uncertainty resulting from wavelength calibration could be 2 km~s$^{-1}$. Thus, an accurate estimation of the inclination angle may not be achieved using this method. In the future we plan to analyze datasets taken around the disk center and use magnetic field extrapolation to derive the inclination angles of the field lines \citep[e.g.,][]{Jess2016} at light bridges. Such an analysis should tell us whether the longer period of the light bridge oscillation is related to the inclination of magnetic field lines or not. It is also possible that two consecutive shocks merge into one shock, as the dominant period of the light bridge oscillations ($\sim$6.5 minutes) appears to be twice as large as that of the umbral oscillations ($\sim$3 minutes). Such shock merging has recently been detected in a pore by \cite{Chae2015}. However, there appears to be no shock merging from Figure~\ref{fig.8}. A third possibility is the interference of two running waves \citep{Su2016}. But from our observation there seems to be no oppositely propagating waves. 

In conclusion, our observations suggest that the persistent surge-like oscillations or light walls above light bridges are caused by shocks. P-mode waves propagate upward along inclined magnetic field lines around the light bridge from the photosphere. They steepen into acoustic shock waves in the chromosphere and lift up the chromospheric materials periodically. We have demonstrated that this scenario has no difficulty in explaining all observational results. However, we do not rule out the contribution of reconnection driven jets to the light bridge oscillations. Magnetic reconnections between emerging fluxes and the sunspot fields can also launch plasma jets/surges at some locations intermittently. But these reconnection driven surges are just occasionally superimposed on the persistent surges produced by shocked p-mode waves. From the literature it appears that most authors did not realize the existence of these two types of surges, probably due to the relatively low quality and/or low resolution of previous observations. Our most recent work has unambiguously revealed both shock-generated and reconnection-related surges in a single observation. The properties of these two types of surge-like activities in this new observation will be discussed in detail in an upcoming paper \citep{Tian2017}. 

We realize that the modulation of p-mode waves may lead to recurrent magnetic reconnection which will produce repetitive jets \citep[e.g.,][]{Chen2006}. However, in order to explain the coordinated behaviors of the surges/jets over the entire light bridge, field lines at different locations on the light bridge need to be reconnected at the same time. On the real Sun a small difference of the magnetic structures between different locations on the light bridge may not lead to the occurrence of reconnection at different locations at the same time, though the p-mode modulation can be coherent over the entire LB. In addition, in our observation the surge speeds are likely lower than the Alfv\'en speed of the chromosphere, which also may not support this scenario. Three-dimensional numerical simulations may be needed to investigate whether p-mode modulated reconnection can cause the coordinated behaviors or not.

\begin{acknowledgements}
IRIS is a NASA small explorer mission developed and operated by LMSAL with mission operations executed at NASA Ames Research center and major contributions to downlink communications funded by ESA and the Norwegian Space Centre. This work is supported by NSFC grants 41574166 and 41574168, the Recruitment Program of Global Experts of China, and the Max Planck Partner Group program. We thank Dr. Shuhong Yang, Prof. Pengfei Chen, Prof. Valery Nakariakov and the anonymous reviewer for helpful discussion and constructive suggestions. 
\end{acknowledgements}

\end{document}